\documentclass[openacc]{rsproca_new}
\usepackage[numbers,sort&compress]{natbib}

\usepackage{xcolor}
\definecolor{revgreen}{RGB}{0,120,0}

\newcommand{\rev}[1]{#1}

\titlehead{Research}

\begin{document}

\title{Chromospheric dynamics and turbulence regulate the solar FIP effect}

\author{
Andy S.H. To$^{1}$, J. Martin Laming$^{2}$, Jeffrey Reep$^{3}$, and Adam J. Finley$^{1}$
}

\address{$^{1}$European Space Agency (ESA), European Space Research and Technology Centre (ESTEC), Keplerlaan 1, 2201 AZ Noordwijk, the Netherlands\\
$^{2}$Space Science Division, Code 7684, Naval Research Laboratory, Washington DC 20375, USA\\
$^{3}$Institute for Astronomy, University of Hawai`i, Honolulu, HI 96822, USA
}

\subject{astrophysics, solar physics, stars, spectroscopy}


\corres{Andy S. H. To\\
\email{andyshto.work@gmail.com}}

\begin{fmtext}
\begin{abstract}
Elemental abundance variations in the solar corona, commonly characterised by First Ionisation Potential (FIP) bias, provide crucial diagnostics of chromospheric processes. The ponderomotive force model, which attributes fractionation to Alfvén wave propagation, has successfully reproduced observed abundance and fractionation patterns in various solar features. However, existing theoretical implementations rely on a static quiet Sun chromosphere, leaving the influence of chromospheric dynamics largely unexplored. We address this limitation by combining hydrodynamic simulations from \texttt{HYDRAD} with ponderomotive force calculations through \texttt{FIPpy}, a new open-source code. Comparing predictions between an initial VAL-C chromosphere and a heated chromosphere following impulsive nanoflare-like events, \rev{we show that the ponderomotive force model remains consistent under dynamic chromospheric conditions, while stronger changes in fractionation behaviour arise from variations in acoustic flux and turbulence.} Most significantly, when acoustic wave flux drops below $\sim5\times10^6$ erg cm$^{-2}$ s$^{-1}$, mass-dependent thermal velocities dominate the fractionation process, producing counterintuitive patterns where Fe exceeds Ca in FIP bias, while high-FIP Ar shows fractionation. We demonstrate that any source of chromospheric turbulence will act to suppress fractionation. For flares, our results predict that the increased turbulence will suppress FIP bias, potentially explaining the observed abundance variations during flares. These findings suggest that coronal abundances and composition encode a sensitive balance between dominant mechanisms, determined by the ratio of ponderomotive acceleration to turbulent velocity.
\end{abstract}

\rsbreak

\section{Introduction}
Elemental abundance variations in the corona serve as markers for physical processes occurring in the chromosphere and transition region. Currently, the most widely adopted theoretical framework for understanding these variations is the ponderomotive force model, which describes fractionation as an effect of Alfv\'{e}n waves propagating through the solar atmosphere~\cite{Laming2009Apr,Laming2015Sep,Laming2021Mar, Laming2023ApJ...951...86L}. The model has successfully reproduced fractionation patterns in various features of the Sun, including fast and slow solar wind~\cite[e.g.,][]{Brooks2010Dec, Brooks2015,Abbo2016SSRv..201...55A, Laming2019ApJ...879..124L, Reville2021FrASS...8....2R}, active regions~\cite[e.g.,][]{Baker2013Nov, Laming2015Sep,Mihailescu2022ApJ...933..245M,Testa2023ApJ...944..117T, Long2024ApJ...965...63L}, and the inverse-FIP effect~\cite[e.g.,][]{Baker2024ApJ...970...39B,Laming2021Mar,Ng2024ApJ...972..123N}, where elements are depleted in the solar corona, reproducing M-dwarf like composition patterns~\cite{Testa2015RSPTA.37340259T,Laming2019ApJ...879..124L,Seli2022A&A...659A...3S}. 

Despite these successes, a critical limitation exists: current theoretical implementations rely predominantly on static quiet Sun chromospheric structures~\cite{Avrett2008ApJS..175..229A}. This limitation has become increasingly apparent as observations reveal substantial temporal and spatial variations in First Ionisation Potential (FIP) bias across different solar conditions~\cite[e.g.,][]{Doschek2018ApJ...853..178D,Baker2019ApJ...875...35B,Mondal2021ApJ...920....4M,To2021ApJ...911...86T,Mihailescu2023ApJ...959...72M,Suarez2023ApJ...957...14S,Baker2024ApJ...970...39B,Ng2024ApJ...972..123N,To2024A&A...691A..95T, Brooks2024ApJ...962..105B,Mondal2025arXiv251002102M}. Observational evidence particularly demonstrates that FIP bias is highly sensitive to local chromospheric conditions~\cite{To2021ApJ...911...86T,Brooks2021SciA....7...68B,To2023ApJ...948..121T,Mihailescu2023ApJ...959...72M, Yardley2024ApJ...976..152Y}.

Recent modelling efforts have begun addressing this gap from different angles. Pioneering 2.5D magnetohydrodynamic models have explored chromospheric dynamics and their role in the ponderomotive force mechanism \cite{Dahlburg2016ApJ...831..160D, Martinez-Sykora2023ApJ...949..112M}. In parallel, other studies have investigated how time-variable abundances affect coronal loop evolution: time-variable abundances were implemented in the \texttt{EBTEL++} model to understand their effects on cooling timescales~\cite{Barnes2016ApJ...829...31B, Brooks2018ApJ...863..140B,Reep2024ApJ...970L..41R}; by assuming spatially and temporally varying abundance factors, HYDrodynamics and RADiation code~\cite[\texttt{HYDRAD};][]{Bradshaw2003A&A...401..699B,Bradshaw2013ApJ...770...12B} simulations have explored the effect of evaporation on transporting fractionated plasma into coronal loops~\cite{Reep2025arXiv250925695R}, and successfully produced coronal rain in impulsive events due to chromospheric evaporation confining fractionated plasma with a significantly higher radiative cooling rate at the looptop~\cite{Benavitz2025ApJ...992....4B}. Together, these studies have advanced our knowledge of how abundance variations affect loop evolution and radiative cooling. However, \rev{most implementations of the ponderomotive force model assume a static quiet-Sun chromosphere~\cite{Vernazza1981ApJS...45..635V,Avrett2008ApJS..175..229A}. Active region chromospheres are highly dynamic due to impulsive heating and evolving density structure. It is therefore unclear whether the predicted fractionation behaviour remains valid under more realistic chromospheric conditions.}

\rev{In this work, we investigate how dynamic chromospheric conditions influence elemental fractionation within the ponderomotive force framework. We utilise \texttt{HYDRAD} to simulate chromospheric profiles under impulsive heating, and apply these atmospheres as input to the ponderomotive force model to evaluate the resulting fractionation patterns.} To facilitate this investigation, we introduce \texttt{FIPpy}\footnote{The code is available at \url{https://github.com/andyto1234/FIPpy}}, an open-source code that combines hydrodynamic modelling with the ponderomotive force model. \rev{This approach allows us to test and explore how changes in chromospheric conditions regulate the predicted elemental abundance patterns by the ponderomotive force model.}


\section{FIP bias simulation}


\rev{We calculate FIP bias with a post-processing module that uses modelled solar atmospheres from HYDRAD simulations as input. FIPpy directly reads the HYDRAD output files and uses these atmospheric profiles for the Alfv\'{e}n wave and fractionation calculations.} HYDRAD is an open-source field-aligned code that solves the conservation equations for mass, momentum, and energy in a two-fluid plasma confined within a magnetic flux tube. It includes optically thick chromospheric radiation approximated in~\cite{Carlsson2012A&A...539A..39C}, thermal conduction with flux limiting terms, and optically thin radiative losses in the corona, calculated with CHIANTI v10~\cite{Dere1997Oct,DelZanna2021ApJ...909...38D}. The modelling \rev{of the FIP effect} consists of three components: (1) solving Alfv\'{e}n wave transport through the chromosphere, (2) calculating the resulting ponderomotive acceleration, and (3) integrating element-specific fractionation along field lines. Each step is detailed below.

\subsection{Alfv\'{e}n wave transport}

We follow the ponderomotive force formulae developed in previous studies~\cite[e.g.,][]{Laming2009Apr,Laming2015Sep}. First, we solve the Alfv\'{e}n wave transport equations to determine wave energy distribution throughout the chromosphere~\cite{Cranmer2005ApJS..156..265C}. The transport equations for upward and downward propagating waves are expressed using the commonly used Els\"{a}sser variables~\cite[e.g.,][]{Elsasser1950PhRv...79..183E, Dobrowolny1980A&A....83...26D, Cranmer2005ApJS..156..265C}:
\begin{align}\label{equ:transport}
    \frac{\partial z_\pm}{\partial t} + (u_s \mp V_A)\frac{\partial z_\pm}{\partial s} = (u_s \pm V_A)\left(\frac{z_\pm}{4H_\rho} + \frac{z_\mp}{2H_A} \right),
\end{align}
where $z_\pm = \delta v \pm \delta B/\sqrt{4\pi\rho}$ are the Els\"{a}sser variables representing Alfv\'{e}n waves propagating \rev{in opposite directions along the background magnetic field}, $u_s$ is the bulk flow speed, $\rho$ is the density, $V_A$ is the Alfv\'{e}n speed, and $H_\rho$ and $H_A$ are the density and Alfv\'{e}n speed scale heights, defined as 
\begin{equation}\label{equ:scale_heights}
\begin{aligned}
H_\rho &= \rho\left(\frac{\partial \rho}{\partial s}\right)^{-1}, 
&\qquad
H_A &= V_A\left(\frac{\partial V_A}{\partial s}\right)^{-1}.
\end{aligned}
\end{equation}
Using the Els\"{a}sser variable, the Alfv\'{e}n wave perturbation and wave magnetic field can be expressed as
\begin{align}\label{equ:wave_vel} 
    \delta v = \frac{1}{2}(z_+ + z_-), \\
    \frac{\delta B}{\sqrt{4\pi\rho}} = \frac{1}{2}(z_+ - z_-), 
\end{align}

To solve Equation~\ref{equ:transport}, we assume an oscillatory time dependent ansatz of $z_{\pm} = z_{\pm}e^{-iwt}$, where $\omega$ is the wave frequency. This reduces the problem to four coupled ordinary differential equations for the real and imaginary components:
\begin{align}
\frac{\partial z_{\pm,R}}{\partial s} &=
\left(
(u_s \pm V_A)\left(
\frac{z_{\pm,R}}{4H_{\rho}} + \frac{z_{\mp,R}}{2H_A}
\right)
+ \omega z_{\pm,I}
\right) / (u_s \mp V_A),
\\
\frac{\partial z_{\pm,I}}{\partial s} &=
\left(
(u_s \pm V_A)\left(
\frac{z_{\pm,I}}{4H_{\rho}} + \frac{z_{\mp,I}}{2H_A}
\right)
- \omega z_{\pm,R}
\right) / (u_s \mp V_A),
\end{align}
where subscripts R and I denote real and imaginary components, respectively. We solve this system by \rev{assuming harmonic time dependence, the time-dependent problem is then reduced to a spatial boundary-value problem, which is then solved numerically. Appropriate boundary conditions are applied at the $\beta=1$ layer,}  where we assume upcoming acoustic waves convert to Alfv\'{e}n waves by mode or parametric conversion. \rev{Details of the boundary conditions are described in Section~\ref{sect:boundary_conditions}} In the simulation, we prescribe an input Alfv\'{e}n wave amplitude and frequency at this boundary.

\subsection{Ponderomotive force and elemental fractionation}
The ponderomotive acceleration experienced by ions in the presence of Alfv\'{e}n wave gradients can be derived from the \rev{Lagrangian for charged particles interacting with an electromagnetic wave field~\citep[][Section 6.2, from Equation 1 to 5]{Laming2015Sep},} and it is expressed as:
\begin{align}\label{equ:ponderomotive_accel}
    a_{pond.}(s
    ) = \frac{c^2}{4} \frac{d}{ds}\left(\frac{\delta E(s)^2}{B(s)}\right),
\end{align}
where c is the speed of light, $B(s)$ is the magnetic field strength, and the wave electric field perturbation, $\delta E$, is given by
\begin{align}
    \delta E^2 = \frac{B(s)^2}{2c^2}(z^2_+ + z^2_-),
\end{align}
with wave velocity perturbation described in Equation~\ref{equ:wave_vel}.

The fractionation of different elements subject to this ponderomotive acceleration is calculated following \cite[][\rev{detailed in Equation 15--22 in \citealp{Laming2015Sep}}]{Schwadron1999ApJ...521..859S,Laming2015Sep,Reville2021FrASS...8....2R}, using the momentum equations for ions and neutrals with the ponderomotive force acting selectively on ions. This leads to an element-dependent density enhancement:
\begin{align}\label{equ:FIP_bias_integration}
    \frac{\rho_w(s_2)}{\rho_w(s_1)} = \exp \left( 2\int^{s_{2}}_{s_{1}} \frac{\xi_wa_{\rm pond.}\nu_\mathrm{eff}}{\nu_{w,i}v^2_w} ds\right),
\end{align}
\rev{where we are integrating along the loop coordinate from a lower solar atmospheric height, $s_1$, to a higher height, $s_2$}, \rev{\(\xi_w\) is the total ionised fraction of element \(w\), i.e. the fraction of the element that is in any ionised state rather than neutral. Equivalently, $\xi_w$ = 1 - $\xi_{w,\mathrm{n}}$, where $\xi_{w,\mathrm{n}}$ is the fraction of element w in its neutral stage. }\(\nu_\mathrm{eff}\) is the effective collision rate accounting for ion-electron coupling, calculated using
\begin{align}
    \nu_{\text{eff}} =(\nu_{w,i}\times \nu_{w, n})/(\xi_w \times \nu_{w, n} + (1-\xi_w)\times \nu_{w, i}).
\end{align}
\rev{\(\nu_{w,i}\) is 
the ion-neutral collision frequency~\cite{Marsch1995A&A...301..261M, 
Schwadron1999ApJ...521..859S}, defined as $\nu_{w,i} = \nu_{\text{ion-}p} + \nu_{\text{ion-H}}$, and $\nu_{w,n} = \nu_{\text{ion-H}} \times (1+(n_p/n_{\text{H}}))$.} We define the ion-proton elastic collision rates as in \cite{Laming2004Oct}, with
\begin{align}
\nu_{\text{ion-}p}
 &= \frac{3.1\times10^4}{A}
\left(\frac{T}{1\times 10^4}\right)^{-3/2}
\left(\frac{n_p}{1 \times 10^{10}\,}\right), \\
\nu_{\text{ion-H}} &\simeq \frac{9.1 \sigma}{1\times 10^{-15}A}\left(\frac{T}{1\times 10^4}\right) \left( \frac{n_\text{H}}{1\times 10^{10}}\right),
\end{align}
where $A$ is the atomic mass number, $\sigma$ is the scattering cross section in ${\mathrm{cm}^{-2}}$ taken from \cite{Marsch1995A&A...301..261M, Laming2004Oct}, $n_p$ and $n_{\rm H}$ are the number density of proton and hydrogen in ${\rm cm^{-3}}$, respectively. Finally, \(v_w\) is the effective 
speed:
\begin{align}\label{equ:v_w}
    v_w^2 = k_B T/m_w + v^2_\mathrm{turb} + v^2_\mathrm{wave}.
\end{align}

Looking at Equation~\ref{equ:v_w}, \(v_\mathrm{wave}\) captures contributions from parametrically generated slow-mode waves arising from Alfv\'{e}n wave perturbations. In this paper, for \(v_\mathrm{turb}\), we associate this to be the acoustic wave amplitude generated by convection-driven acoustic waves propagating upward with speed, $c_s$, assuming an upward acoustic wave flux ($F_{\rm ac}$) of $10^8~\mathrm{erg~cm^{-2}~s^{-1}}$ that is conserved with height in the WKB approximation. The amplitude of the acoustic wave can then be calculated using
\begin{equation}
    v^2_{\text{turb}} = F_{\text{ac}}/\rho c_s.
\end{equation}
From the two footpoints of the loop,  $v_{\text{turb}}$ is allowed to grow until the amplitude reaches the local sound speed. At this point, the amplitude is capped at the value, similar to the treatment taken in~\cite{Laming2019ApJ...879..124L}, arguing that beyond this point, the excess energy is lost to the wave by radiation and conduction.

The role of turbulent velocities in Equation~\ref{equ:v_w} is crucial for understanding FIP fractionation patterns. Any source of turbulence, whether from e.g., acoustic waves, or other chromospheric dynamics, enters the denominator of the fractionation integral (Equation~\ref{equ:FIP_bias_integration}) through $v^2_w$ \rev{with the $v_{wave}$ term}~\citep[e.g.,][]{Reville2021FrASS...8....2R}. This has a suppressing effect on FIP bias. For a given Alfv\'{e}n wave amplitude, increased turbulence reduces the relative importance of ponderomotive acceleration, $a_{\rm pond.}$, thereby decreasing elemental fractionation. 

\rev{Focusing on $v_{\mathrm{wave}}$, in this study, we consider the parametrically generated slow-mode wave driven by the Alfv\'en wave.} For the fundamental mode ($n=1$), the slow-mode wave velocity can be written as:
\begin{equation}
\delta v_\mathrm{slow} = \frac{-i}{2}\left[\sqrt{\delta v^2 + (H_\rho \omega_s (1-c_s^2/V_A^2))^2} - H_\rho \omega_s (1-c_s^2/V_A^2)\right],
\end{equation}
where $\omega_s = 2\omega_A$ for the fundamental mode, $H_\rho$ is the density scale height, $\delta v$ is the wave velocity perturbation calculated following Equation~\ref{equ:wave_vel}. The magnitude, $|\delta v_\mathrm{slow}|$, contributes to $v_{\rm wave}^2$ in Equation~\ref{equ:v_w}, providing some saturation effects of the FIP bias at high wave amplitudes.

\subsubsection{Ionisation fraction}

\begin{figure}
    \centering
    \includegraphics[width=0.8\linewidth]{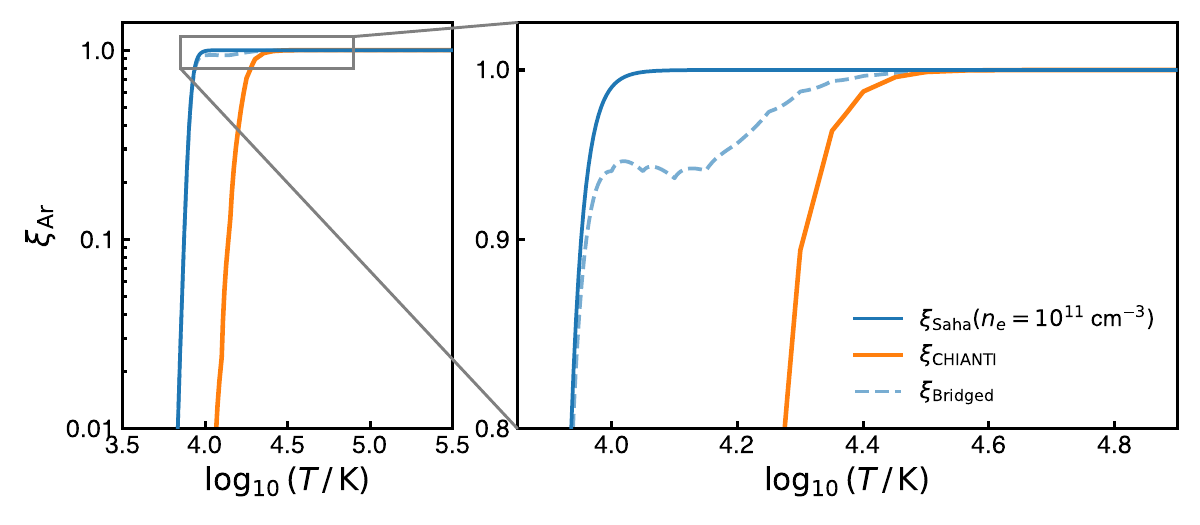}
    \caption{\rev{Plot of the ion fraction of Ar, $\xi_\mathrm{Ar}$, at $n_e=10^{11}~\mathrm{cm^{-3}}$ across temperature to demonstrate the blending between CHIANTI and Saha equation.}
}
    \label{fig:ionisation_fraction_illustration}
\end{figure}

Following Reville et al.~\cite{Reville2021FrASS...8....2R}, \rev{the fraction of the element, $w$, that is in any ionised state rather than neutral}, $\xi_w$, is calculated using a hybrid approach that combines the Saha equation in the chromosphere with CHIANTI atomic database ionisation equilibrium tables~\citep{DelZanna2021ApJ...909...38D} in the corona. \rev{Prior to CHIANTI v11, CHIANTI uses the coronal approximation that assumes ionisation and recombination only due to collisions}, we use the Saha equation below $10^4$~K where radiative processes become important in determining the ionisation balance. \rev{The Saha equation, which depends on both temperature and electron density, is evaluated at the local $(T, n_e)$ at each grid point along the loop.} To ensure smooth continuity between the two regimes, we employ a hyperbolic tangent weighting function in $\log_{10}(T)$ space:
\begin{equation}
f(T) = \frac{1}{2}\left(1 + \tanh\left(\frac{\log_{10}(T) - \log_{10}(T_\mathrm{trans})}{\Delta\log(T)}\right)\right)
\end{equation}
The ionisation fractions are blended in logarithmic space to better capture their exponential variation with temperature:
\begin{equation}
\log_{10}(\xi_w) = (1-f(T))\log_{10}(\xi_\mathrm{Saha}) + f(T)\log_{10}(\xi_\mathrm{CHIANTI})
\end{equation}
\rev{Figure~\ref{fig:ionisation_fraction_illustration} shows the CHIANTI, Saha equation, and the blended ionisation fraction for Ar as an example.} This weight function ensures that: (i) the Saha equation dominates for $T \lesssim 10^4$ K (chromospheric conditions), (ii) CHIANTI data dominates for $T \gtrsim 10^{5}$ K, and (iii) a \rev{relatively} smooth transition occurs through the transition region. For all elements, we use $\log_{10}(T_\mathrm{trans}) = 4.5$ ($\approx$31,600 K) and $\Delta\log(T) = 0.2$ dex. The effect of the interpolation is a relatively gradual transition between the Saha equation's and CHIANTI's ionisation fraction. \rev{However, the Saha and CHIANTI ionisation fractions have different temperature dependences, this blending can introduce small local non-monotonic features in the transition region; these are an artefact of the interpolation rather than a physical decrease in ionisation with increasing temperature.}

\subsubsection{Magnetic field strength}

To evaluate the plasma beta and Alfv\'{e}n speed along the simulated loop, we follow previous studies to model the magnetic field strength as a function of position along the loop coordinate, \rev{$s$}, \rev{only in the \texttt{FIPpy} FIP bias calculation, instead of an input in \texttt{HYDRAD}}~\cite{Russell2013ApJ...765...81R,Reep2016ApJ...818L..20R,Reep2018ApJ...853..101R}. Specifically, we describe the magnetic field strength in gauss as a fifth-order polynomial equation:
\begin{equation}
\begin{aligned}
B(s) ={}& 1031.61 
 - 1.33014\times10^{-5}\,s 
 + 8.48378\times10^{-14}\,s^2 \\
& - 3.03276\times10^{-22}\,s^3 
 + 5.88384\times10^{-31}\,s^4 
 - 4.7167\times10^{-40}\,s^5~[\mathrm{G}]
\end{aligned}
\end{equation}
to interpolate between $\sim$1000~G at the footpoint, and $\sim$100~G at the corona~\cite{Brooks2021ApJ...915L..24B}. In this equation, \(s\) is the coordinate along the loop in cm. The magnetic field strength at the $\beta=1$ layer (where gas pressure equals magnetic pressure) is around $750$~G ($\sim240~\rm km$ above the loop footpoint).

\subsection{Boundary conditions and integration domain}\label{sect:boundary_conditions}

\begin{figure}
    \centering
    \includegraphics[width=0.7\linewidth, trim = 20cm 12cm 20cm 17cm]{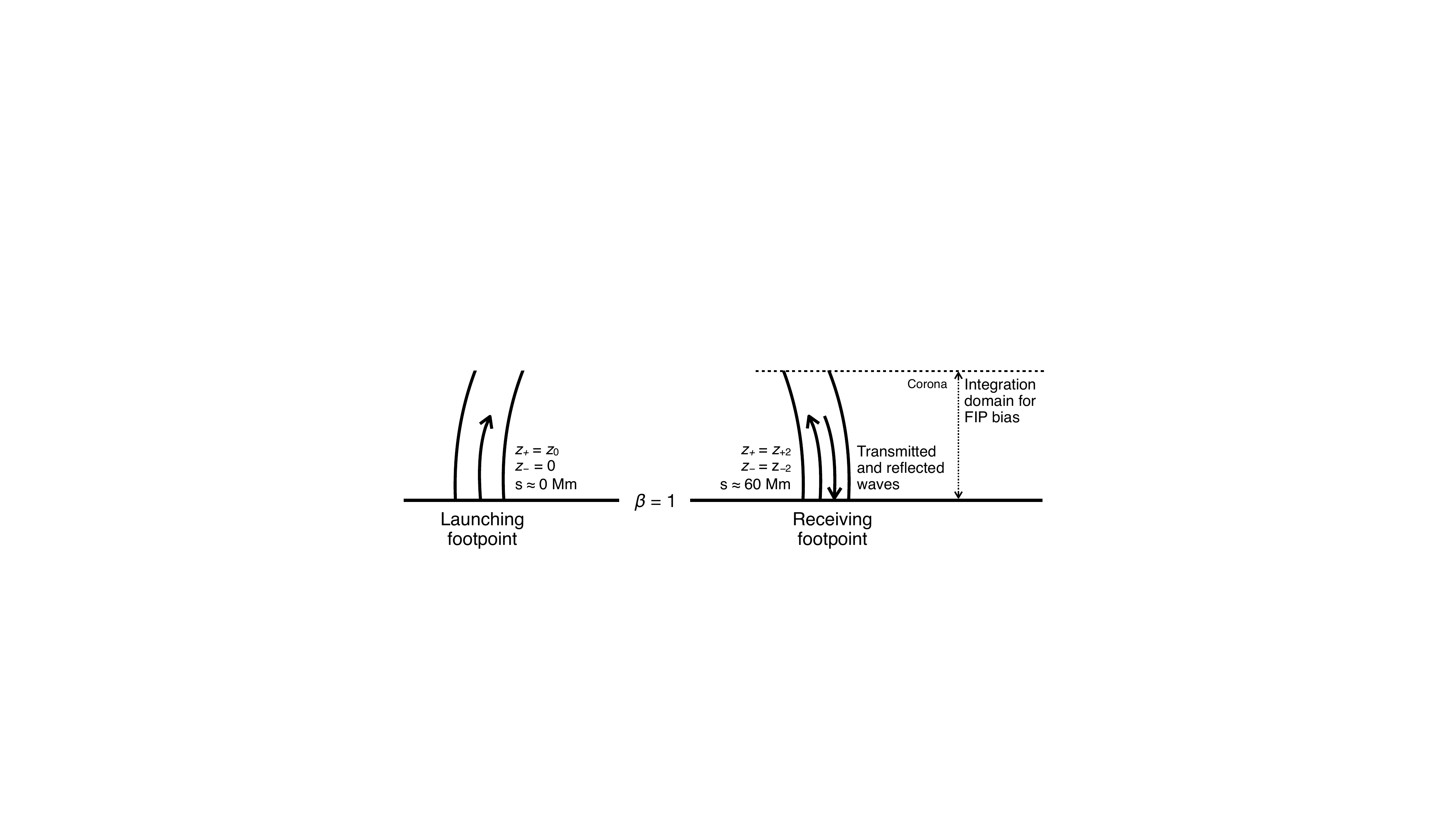}
    \caption{\rev{Schematic of the model setup used for the Alfv\'{e}n wave transport and FIP-bias calculation. Alfv\'{e}n waves are injected at the launching footpoint at the $\beta=1$ layer, and propagate along the loop toward the receiving footpoint, where transmitted and reflected components determine the ponderomotive acceleration, and therefore the predicted FIP bias.}}
    \label{fig:cartoon_ponderomotive_force}
\end{figure}

To model Alfvén wave propagation and fractionation, we treat two chromospheric footpoints of the HYDRAD loop differently, where we have a \textit{launching footpoint} where 
waves are injected, and a \textit{receiving footpoint} where fractionation is evaluated. \rev{Figure~\ref{fig:cartoon_ponderomotive_force} shows a cartoon of the model setup.}  At the launching footpoint, we impose boundary conditions at the \(\beta=1\) 
surface, arguing that most wave mode conversion into Alfv\'{e}n waves occurs around this height, with
\begin{align}
    z_+(s_{\beta=1,\text{launching}}) &= z_0, \\
    z_-(s_{\beta=1,\text{launching}}) &= 0,
\end{align}
where \(z_0\) is the prescribed wave amplitude in km\,s\(^{-1}\). This 
represents upward-propagating Alfv\'{e}n waves with no initial downward component. 
As waves propagate to the \textit{receiving} footpoint, they reflect and generate an opposite component \(z_-\).

The FIP bias (Equation~\ref{equ:FIP_bias_integration}) is evaluated at the \textit{receiving footpoint} by integrating from a reference height \rev{$s_1\approx 60~\mathrm{Mm}$} in the lower chromosphere, where FIP bias = 1 for all elements, to \rev{\(s_2\)} in the low corona (roughly 59.5~Mm to 57.5~Mm in this paper), where ionisation is complete and fractionation ceases. This integration captures the cumulative effect of both upward-propagating (transmitted) and downward-propagating (reflected) waves. Normalising to oxygen yields element abundance enhancements directly comparable to solar wind observations:
\begin{equation}
\mathrm{FIP~bias}_w = \frac{\rho_w(s_1)/\rho_w(s_0)}
{\rho_{\mathrm{oxygen}}(s_1)/\rho_{\mathrm{oxygen}}(s_0)}.
\end{equation}
\rev{Since the absolute amplitude of Alfv\'{e}n waves in coronal loops is poorly constrained observationally, we normalise the wave amplitude such that Ca/O $\sim4$ and focus on changes in the relative fractionation pattern.}

\section{Simulations}

To investigate how chromospheric structure affects FIP fractionation under the ponderomotive force framework, we solve the Alfv\'{e}n wave transport equation and evaluate the predicted FIP bias in two scenarios simulated by HYDRAD: (1) the initial quiet Sun chromosphere \cite[VAL-C; ][]{Vernazza1981ApJS...45..635V}, and (2) a heated chromosphere following four impulsive heating events. \rev{In doing so, we address two related questions. First, by comparing the elemental abundances predicted by a quiet-Sun like VAL-C atmosphere and a dynamically heated chromosphere, we test whether the ponderomotive force model remains qualitatively consistent in different chromospheres. Second, we investigate that within the ponderomotive force model, which chromospheric terms (including acoustic wave flux and additional turbulence), most strongly regulate the resulting fractionation pattern.}

\subsection{Setup and parameters}

Both simulations employ a semicircular coronal loop of total length $s = 2L = 60$~Mm, with the initial transition region \rev{(defined here as the location where the heat flux changes from a sink in the corona to a source in the transition region)} positioned at $s = 2.26$~Mm, and $60-2.26=57.74$~Mm. \rev{The initial VAL-C atmosphere is solved independently in \texttt{HYDRAD}, following the implementation described in \cite{Reep2019ApJ...871...18R}. This yields low-coronal properties of $T \approx 1 \times 10^5$~K and $n_e \approx 10^9$~cm$^{-3}$} at $s = 57.5$~Mm. Radiative cooling is computed using the 15 most abundant solar elements in the corona~\cite{Asplund2009ARA&A..47..481A}, with optically thick radiation enabled in the chromosphere, which includes contributions from H, Ca, and Mg.

We select Alfv\'{e}n wave parameters that satisfy the resonance condition, where
\begin{equation}
    f_{\rm resonance} \simeq \frac{<V_{\rm A}>}{2L},
\end{equation}
and roughly equal Alfv\'{e}n wave flux for upward and downward propagating waves throughout the loop. For both VAL-C and heated chromospheres, we adjust the wave amplitude such that the predicted Ca/O FIP bias reaches $\sim4$, \rev{consistent with the typical reported elemental abundances in coronal and solar wind, relative to a high-FIP reference element~\cite[e.g.,][]{Raymond2001AIPC..598...49R,Brooks2010Dec,Abbo2016SSRv..201...55A,Heber2021ApJ...907...15H, Alterman2025A&A...700A..23A}, we then focus on the relative ordering and element-to-element variation in FIP bias under different chromospheric conditions.}

\begin{figure*}
    \centering
    \includegraphics[width=\linewidth]{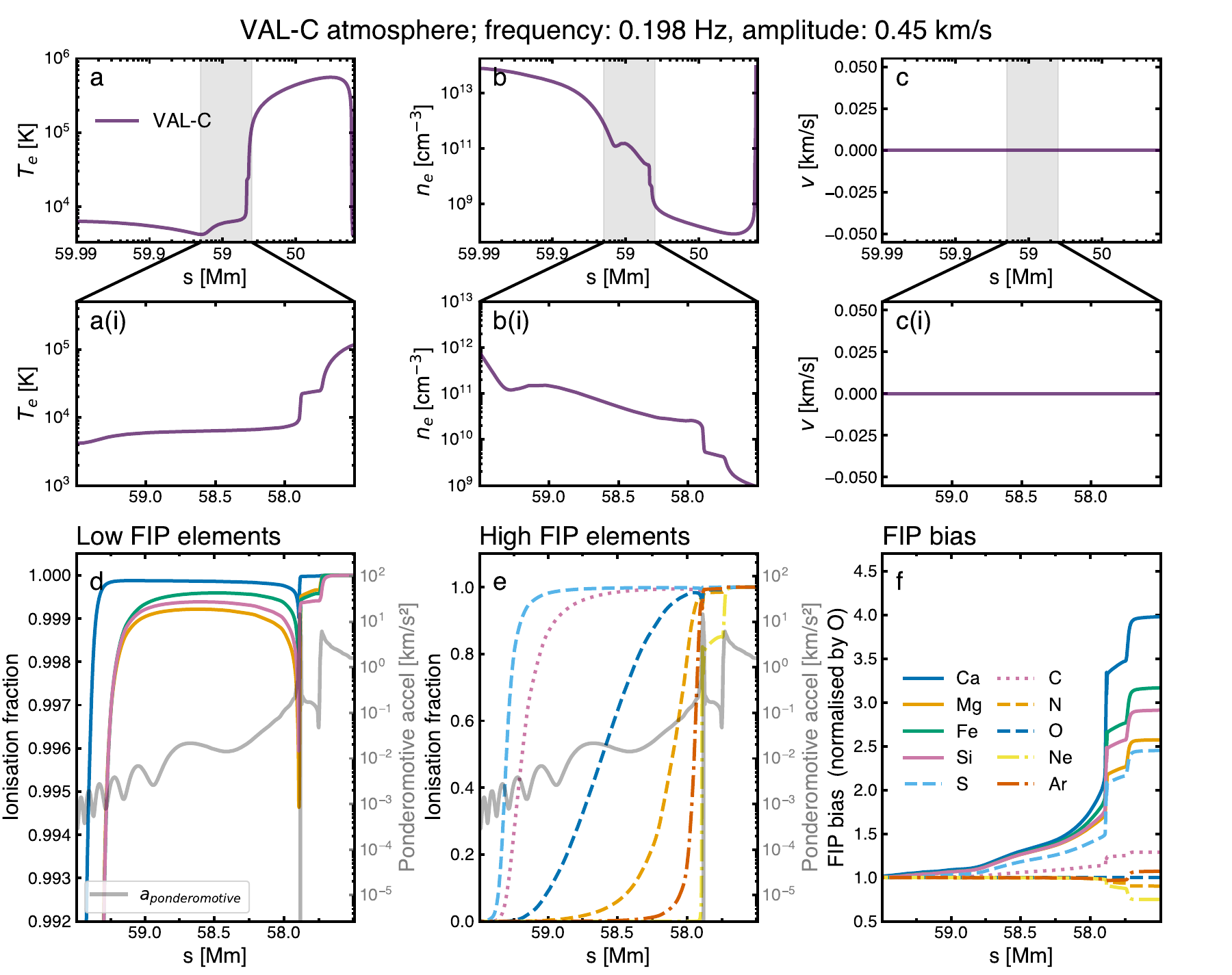}
    \caption{\rev{HYDRAD chromospheric simulation using the VAL-C atmosphere, and the predicted FIP bias values. X-axis are flipped and in log scale to show the receiving end of the footpoint.} The top row shows $T_{e}$, $n_{e}$, and $v$. Grey shaded area indicates the region of interest in the middle row. The bottom three panels illustrate (d) ionisation fractions of low‑FIP elements (Ca, Mg, Fe, Si), (e) ionisation fractions of high‑FIP elements (O, S, Ar, C), and (f) the resulting FIP bias normalised by O, together with the ponderomotive acceleration profile (grey dashed line; panel g and h). The simulation corresponds to an initial VAL‑C chromosphere before any heating, with an Alfv\'{e}n wave frequency = $0.198~\text{Hz}$, and amplitude = $0.45~\text{km}~\text{s}^{-1}$.}
    \label{fig:0s}
\end{figure*}

\subsection{\rev{VAL-C atmosphere}}

Figure~\ref{fig:0s} presents results for the initial VAL-C model. At these quiet Sun conditions, injecting wave with frequency = 0.198 Hz, and amplitude = 0.45 km\,s\(^{-1}\) (corresponding to $\delta B_0/B_0 \approx 0.03$), the ponderomotive acceleration strongly peaks at the chromosphere/transition region interface (grey dashed line, panel g and h), reaching $\sim 10^2$ km\,s$^{-2}$ before dropping sharply into the corona.

\begin{figure*}
    \centering
    \includegraphics[width=\linewidth]{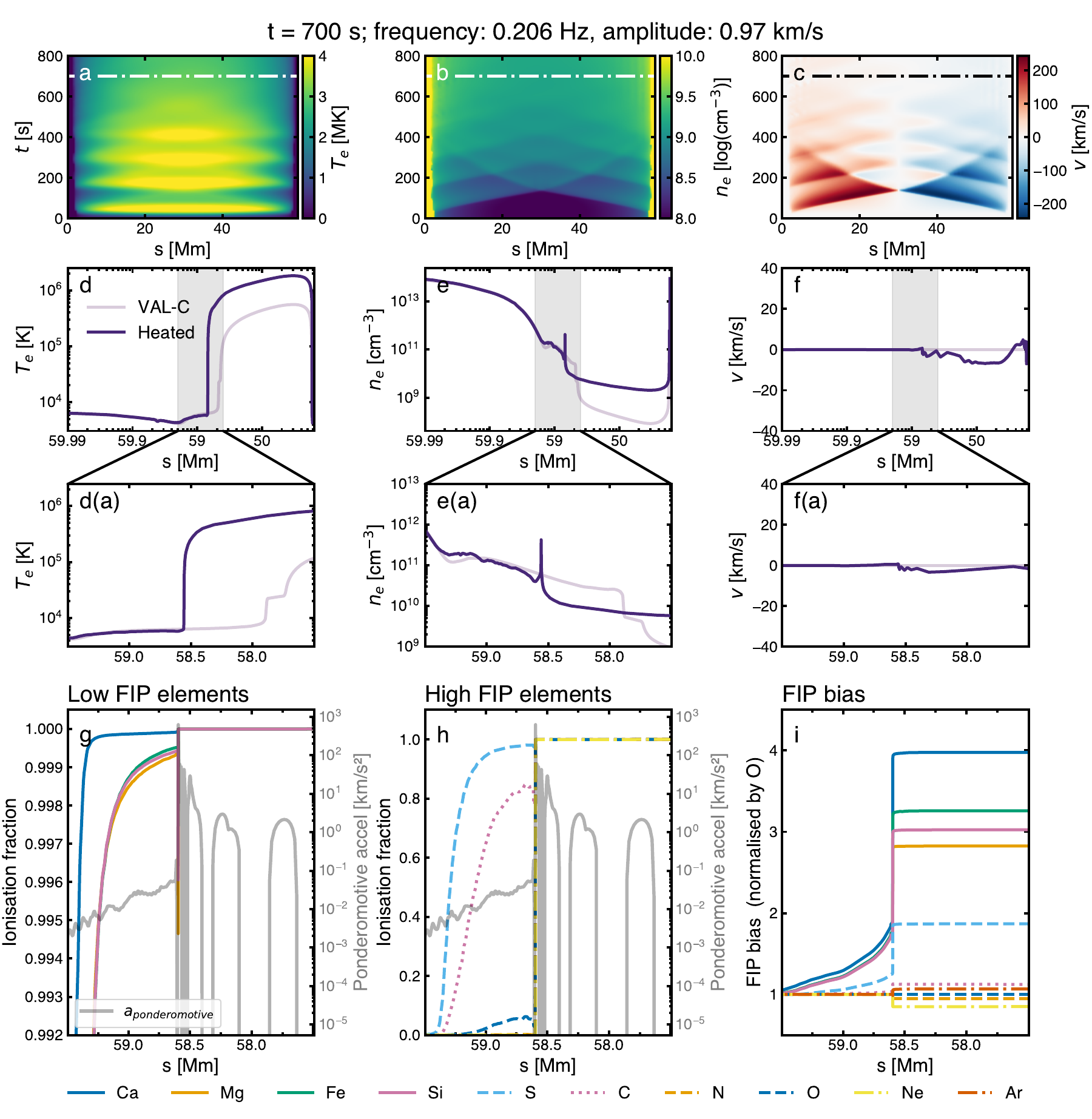}
    \caption{HYDRAD chromospheric simulation after heating (t = 700 s, \rev{indicated by the dashed horizontal line in the first row}) and the resulting FIP bias pattern. Panels are arranged similarly as in Figure~\ref{fig:0s}. The loop underwent 4 heating events, resulting in a hotter, and denser transition region, with coronal peak temperature $\sim2.5~\text{MK}$. \rev{Previous VAL-C atmosphere is shown as the faint purple line in the middle rows (panel d--f), with the vibrant purple line showing the heated atmosphere.} Compared to the VAL-C chromosphere, the altered ionisation structure, the heated chromosphere shows different ponderomotive acceleration profiles, influencing the predicted FIP bias. Alfv\'{e}n‑wave frequency = $0.206~\text{Hz}$, amplitude = $0.97~\text{km}~\text{s}^{-1}$.}
    \label{fig:700s}
\end{figure*}

The resulting acceleration and fractionation pattern is similar to FIP bias results from previous studies~\cite[e.g.,][]{Laming2009Apr,Laming2015Sep} that modelled FIP bias using Avrett $\&$ Loeser's chromosphere~\cite{Avrett2008ApJS..175..229A}. \rev{By design, we normalised the Alfv\'{e}n wave amplitude so that Ca/O reaches a value of $\sim4$, we therefore focus on the relative fractionation pattern among the remaining elements.} Low-FIP elements (Mg, Fe, Si) show pronounced enhancements relative to high-FIP elements (C, N, and Ar), with abundance ratios of $\sim 3$ for Fe, Si, and Mg, and some fractionation of S (\(\sim 2.5\)). High-FIP elements Ar and O show no enhancement, remaining at photospheric abundances. The ionisation fractions (panels d and e) exhibit sharp transitions at the chromosphere/transition region boundary (\(s \approx 58\) Mm), reflecting the steep temperature gradient from \(\sim 10^4\) K to \(\sim 10^5\) K, as well as effect due to interpolation artifacts between the CHIANTI and Saha models. This effect is nonetheless very minor for low FIP elements.


\subsection{Heated chromosphere}

Figure~\ref{fig:700s} shows the configuration after four consecutive heating events, each delivering \(0.54 \, \text{erg} \, \text{cm}^{-3}\) uniformly across the loop, spaced 90 s apart. The FIP bias is evaluated at \(t = 700\)~s once the bulk velocity has relaxed. Top row of Figure~\ref{fig:700s} shows a 2D representation of the heatings with its effects on electron temperature, density and bulk flow velocity. 

The heating produces a very different chromospheric-coronal profile: the final loop profile is indicated by the yellow curve in the second and third rows of Figure~\ref{fig:700s}. After 4 heating events, the loop reaches $\sim 2.5$ MK, substantially hotter than the VAL-C case $\sim~0.5$~MK, while the transition region becomes denser and more extended, with the lowest electron density in the corona at $n_e\sim2\times10^{9}~\mathrm{cm^{-3}}$. \rev{Here, we identify the transition region as the narrow layer where the temperature rises rapidly from chromospheric to coronal values at $s\approx58.4$~Mm.} In the resonant case with a wave frequency of $0.206~\text{Hz}$ and amplitude of $0.97~\text{km}~\text{s}^{-1}$ (adjusted to maintain resonance in the altered loop structure; corresponding to $\delta B_0/B_0 \approx 0.06$). \rev{The slight shift in frequency reflects the heating} modifying both the Alfv\'{e}n-speed profile and the location of the transition region. The transition region acts as the main reflection layer and thus sets the effective length of the lower-atmospheric resonant cavity. Ponderomotive acceleration remains strongly peaked in the chromosphere at $\sim 10^7~\text{cm}~\mathrm{s}^{-2}$. 

The increased density in the \rev{0.5--1.5 Mm} range of the heated chromosphere \rev{from the footpoint boundary} suppresses ionisation fractions obtained from the Saha equation, leading to overall lower ionisation fractions for all elements in the chromosphere. Consequently, this different chromospheric/transition region profile results in a \rev{very similar yet} subtly different predicted FIP bias. \rev{Figure~\ref{fig:by_element_fip_bias_comparison} shows the predicted FIP bias patterns and their differences between the two atmospheres.} Compared to the VAL-C chromosphere, the heated chromosphere shows \rev{similarly enhanced fractionation for low-FIP elements: Fe remains highly fractionated with FIP bias of $\sim$3.3}, while Si and Mg reach FIP bias of $\sim$3.0. The intermediate-FIP element S shows a more modest fractionation \rev{with the largest difference between the two atmospheres} \rev{($\sim$1.8, difference of 0.6)}, while high-FIP elements such as Ar and N remain unfractionated, consistent with expectations.

\begin{figure}
    \centering
    \includegraphics[width=0.9\linewidth]{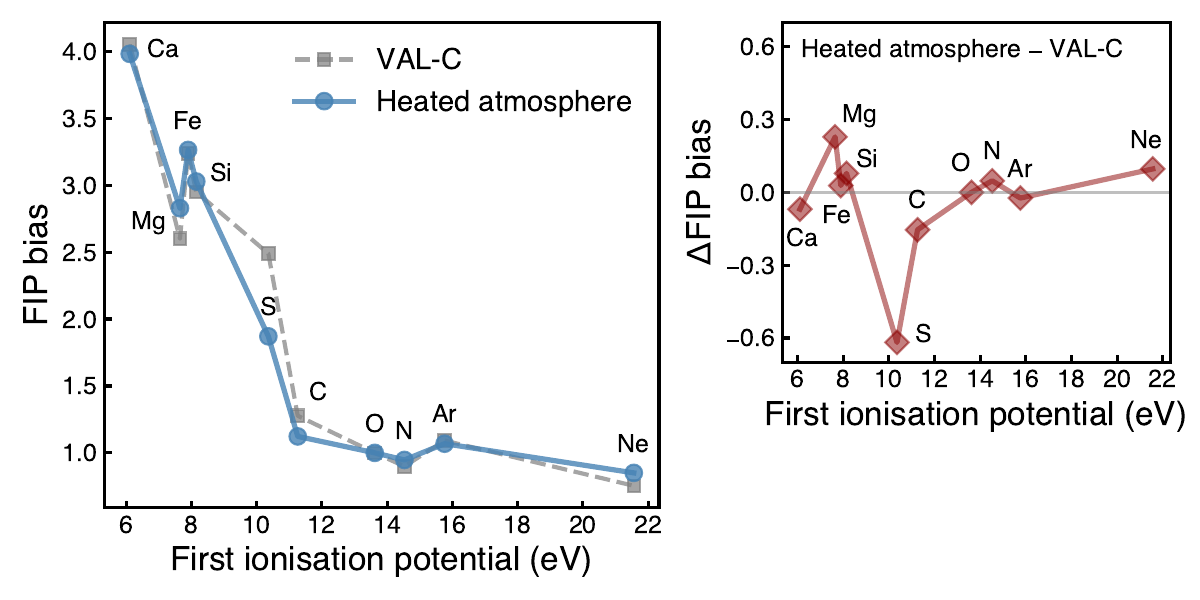}
    \caption{\rev{Variation of the predicted FIP bias between the VAL-C and heated atmospheres, evaluated at s = 57.5~Mm. Right panel shows the FIP bias difference between the two atmospheres. FIP biases are normalised by O.}}
    \label{fig:by_element_fip_bias_comparison}
\end{figure}

\subsection{Acoustic wave flux}

Having established the predicted FIP bias for an active region-like loop, we now investigate how acoustic wave flux, a key source in Equation~\ref{equ:v_w}, regulates fractionation patterns under this atmospheric structure. The code developed here allows systematic exploration of how this critical chromospheric parameter determines elemental abundance patterns.

\begin{figure}
    \centering
    \includegraphics[width=\linewidth]{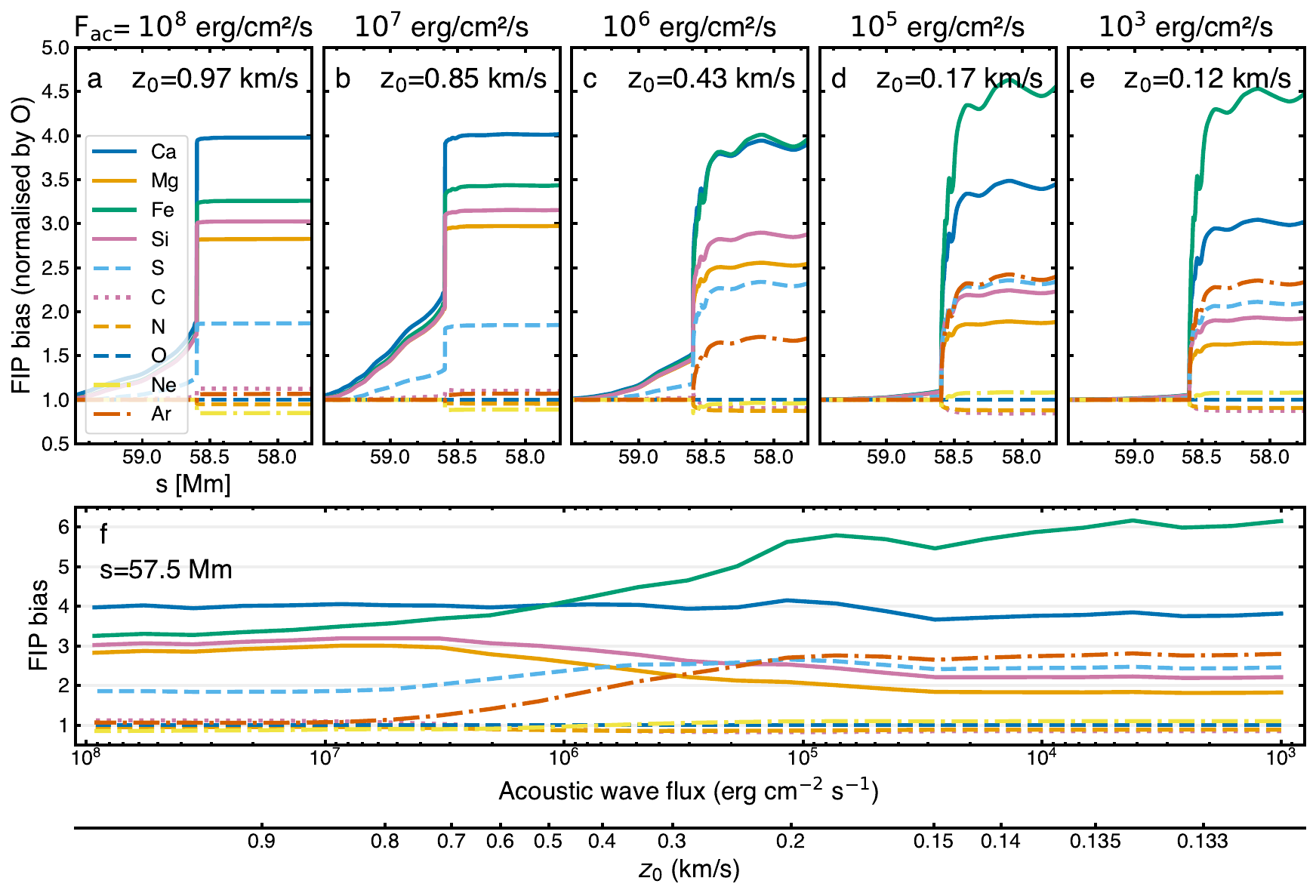}
    \caption{Variation of the predicted FIP bias with acoustic wave flux ($F_{\text{ac}}$) using the heated chromosphere, with the range of acoustic wave flux taken from~\cite{Finley2022A&A...665A.118F}. The top row shows the predicted FIP bias at decreasing acoustic wave fluxes from $10^8~\mathrm{erg~cm^{-2}~s^{-1}}$ to $10^3~\mathrm{erg~cm^{-2}~s^{-1}}$. \rev{Here, we adjusted the input wave flux to show FIP biases that are roughly consistent with solar observations for a comparison of each element's behaviour}. Bottom row shows how FIP bias varies across all elements at $s=57.5~\mathrm{Mm}$ when Ca/O is fixed at $\sim 4$ for each acoustic flux value to better show the comparative behaviour of elements.
}
    \label{fig:acoustic_investigation}
\end{figure}

3D MHD simulations indicate that acoustic wave fluxes in the photosphere vary considerably, spanning \(10^3-10^8~\mathrm{erg~cm^{-2}~s^{-1}}\)~\citep[typical upward acoustic fluxes of order $10^6-10^8~\mathrm{erg~cm^{-2}~s^{-1}}$; ][]{Heggland2011ApJ...743..142H, Finley2022A&A...665A.118F}. To explore how the predicted FIP bias responds to this variation, we use the heated chromosphere with different acoustic flux values. The top row of Figure~\ref{fig:acoustic_investigation} shows the predicted FIP bias patterns across a range of acoustic wave fluxes, where we adjusted the Alfv\'{e}n waves amplitudes to reproduce solar-like fractionation, with a \rev{Ca/O} FIP bias of $\sim4$. \rev{The bottom row} shows the overall fractionation pattern across acoustic wave flux, varying \rev{wave amplitude, $z_0$}, by only fixing the Ca/O FIP bias to be $\sim4$.

It can be seen that weaker acoustic wave flux corresponds to a weaker launching Alfv\'{e}n wave amplitude (across panel a to e) to create similar degrees of fractionation. Simultaneously, the FIP bias pattern changes significantly as the wave flux decreases. For fluxes above $10^7\geq\mathrm{erg~cm^{-2}~s^{-1}}$, \rev{the comparative pattern of FIP bias across different elements remain roughly the same.} However, \rev{as $F_\text{ac}$ continues to decrease, Fe now} develops FIP bias comparable to, or even greater than Ca. \rev{High FIP elements including S, and Ar} also exhibit increasingly strong fractionation. The extreme case of $F_\mathrm{ac} = 10^3~\mathrm{erg~cm^{-2}~s^{-1}}$ is particularly noteworthy. Now, it can be seen that Fe shows the highest FIP bias, \rev{while S, and Ar} displays unexpectedly strong fractionation, with FIP bias exceeding even some low FIP elements such as Mg and Si. Since sunspot umbrae likely exhibit a reduced upward acoustic wave flux, it is plausible that reduced acoustic wave flux might help explain localised inverse-FIP abundance~\citep[e.g.,][]{Doschek2017ApJ...844...52D,Baker2019ApJ...875...35B,Baker2024ApJ...970...39B}.

\begin{figure}
    \centering
    \includegraphics[width=\linewidth]{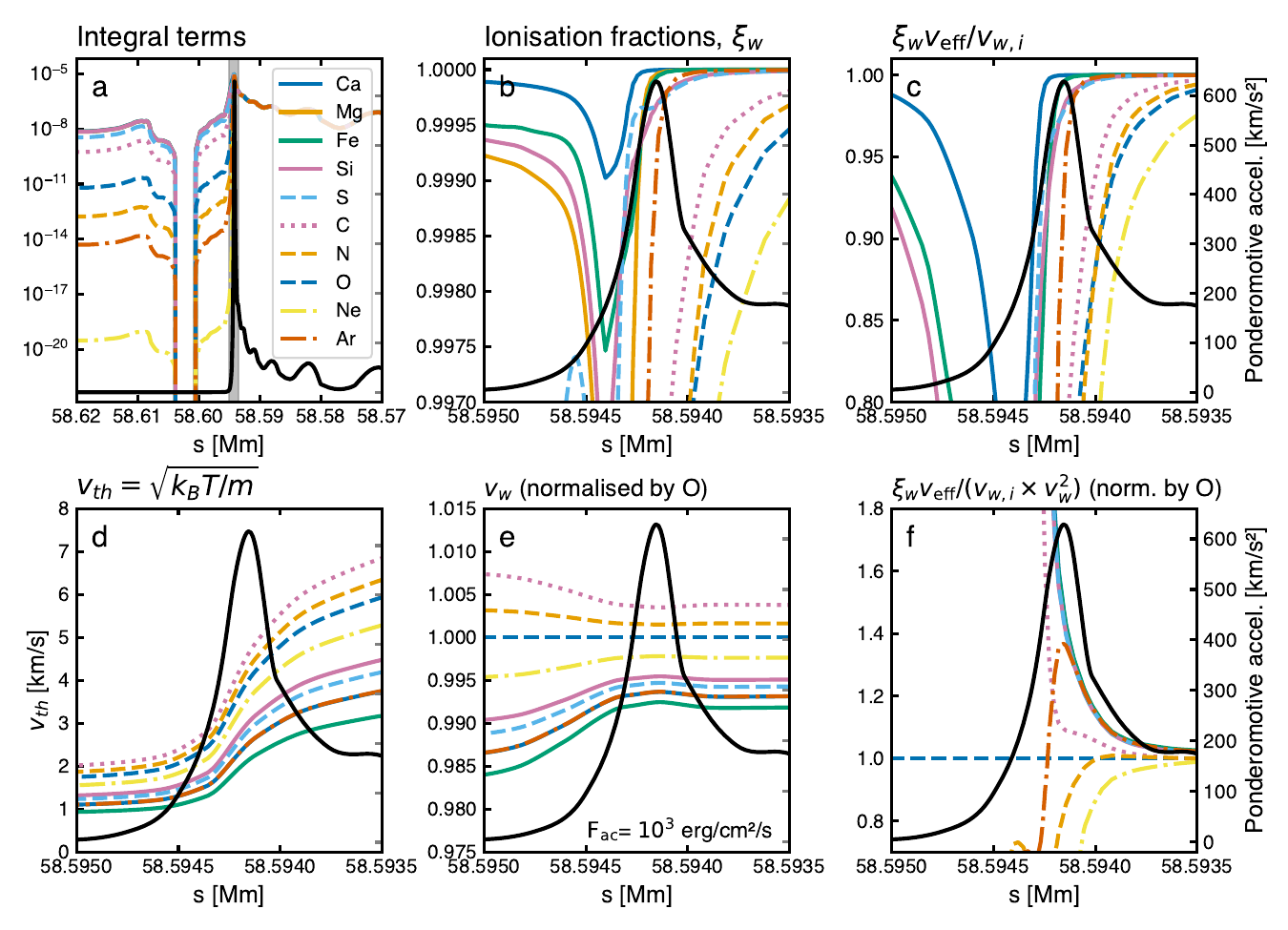}
    \caption{Investigation of the integrand in Equation~\ref{equ:FIP_bias_integration} for a weak acoustic wave flux of $10^3~\mathrm{erg~cm^{-2}~s^{-1}}$. \rev{Panels b--f show the region highlighted by the grey shaded area in panel a. The solid black line denotes the ponderomotive acceleration.} Panels a--e show the individual terms, or their combined effects, while panel f shows how the integral term is redistributed to produce a mass-dependent behaviour.}
        \label{fig:integral_investigation}
\end{figure}

To understand this behaviour, Figure~\ref{fig:integral_investigation} dissects the integral term in Equation~\ref{equ:FIP_bias_integration} under the weakest acoustic wave flux condition ($10^3~\mathrm{erg~cm^{-2}~s^{-1}}$). The black solid line indicates the ponderomotive acceleration. From left to right, top to bottom, panel (a) shows the final integral values that determine FIP bias. Panels (b--c) demonstrate that the numerator terms (ionisation fractions \(\xi_w\) and collision frequency ratios multiplied by ionisation fractions \(\xi_w v_{\rm eff}/v_{s,i}\)) around the peak of the ponderomotive acceleration. From panel (c), it can be seen that around the ponderomotive acceleration peak, elements are ordered roughly according to their FIP. However, as we add the thermal speed term, $v_{\rm th}$, and $v_{w}$ ($k_BT/m_w + v_{\rm turb}^2 + v_{\rm wave}^2$) in the denominator of Equation~\ref{equ:FIP_bias_integration}, this picture changes considerably. Panel (d) shows thermal speeds alone. This term is inherently mass-dependent. Heavier elements have systematically lower thermal speeds in local thermodynamic equilibrium (LTE). For instance, as $a_{\rm pond.}$ peaks, Fe (mass 56 amu) has \(v_{\rm th} \approx 1.5\) km s\(^{-1}\), while C (mass 12 amu) reaches \(v_{\rm th} \approx 3.5\) km s\(^{-1}\), and Ar (mass 40 amu) leans towards the heavier element side, at \(\approx 1.8\) km s\(^{-1}\). When acoustic wave contributions to \(v_{\rm turb}\) are weak, this thermal term dominates the denominator.

Panel (e) shows the combined effective speed \(v_w\) including all other contributions, including slow mode and acoustic waves. At this low acoustic flux, the thermal speed differences remain significant, preserving mass-dependent ordering. Since the fractionation integral depends on \(1/v_w^2\), heavier elements with lower \(v_w\) experience stronger fractionation. This explains why Fe with both low FIP (7.90 eV) and high mass (56 amu) achieves the strongest enhancement. Similarly, Ar's high mass (40 amu) partially compensates for its high FIP (15.76 eV), producing modest fractionation that would be absent under simple FIP ordering.

Panel (f) shows the final normalized integral term, illustrating how this mass dependence manifests in the predicted FIP bias, with mass dependent effect now fully propagated to the FIP bias calculation. This results in a complex pattern where both FIP and mass contribute to determining elemental enhancements, \rev{could be} testable in observations of regions with suppressed acoustic wave flux.

\subsection{Additional turbulence}

\begin{figure}
    \centering
    \includegraphics[width=\linewidth]{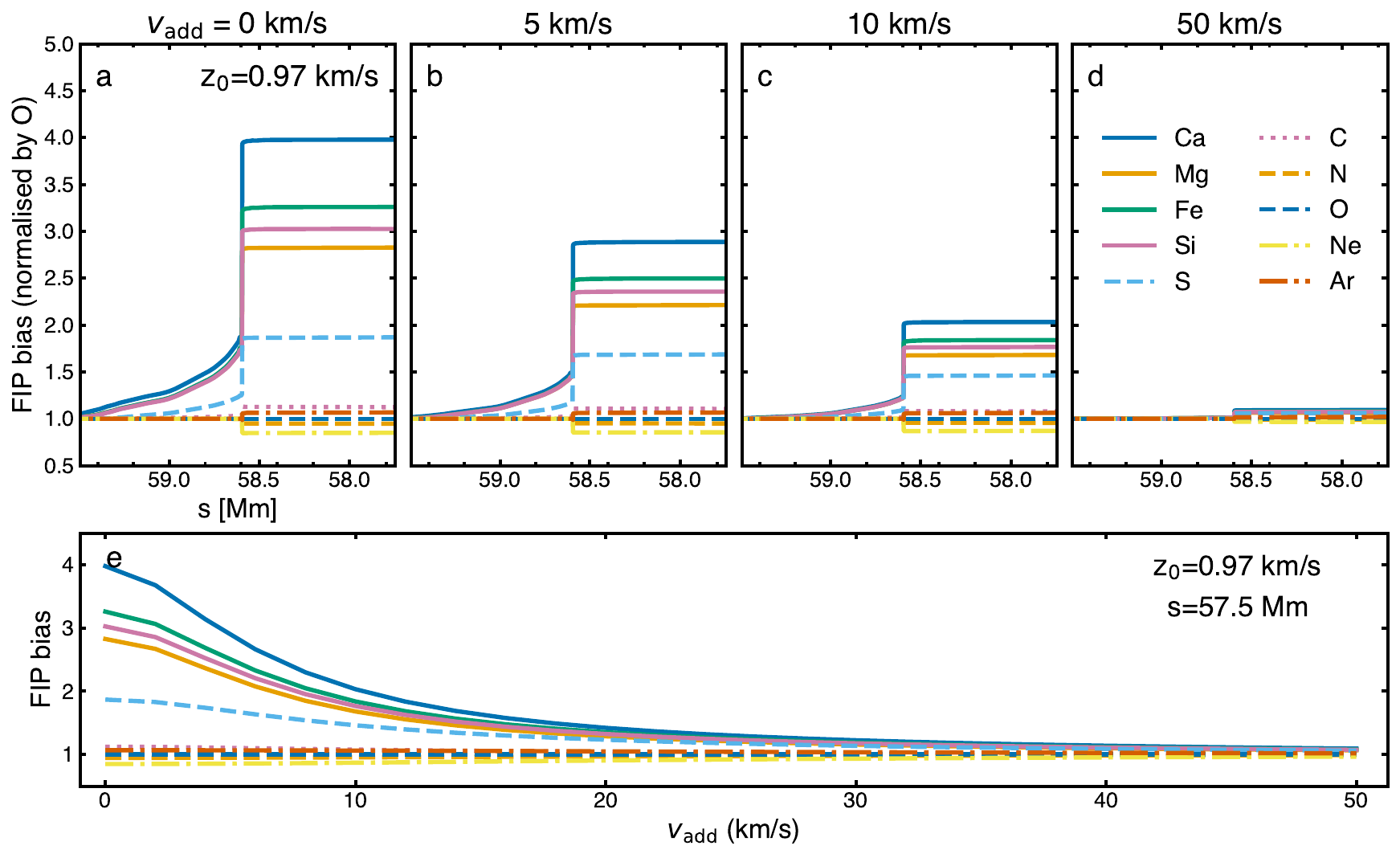}
    \caption{Variation of the predicted FIP bias with an artificially added turbulence ($v_{add}$) at t=700~s at different values, with $F_{\rm ac} = 10^8~\rm erg~cm^{-2}~s^{-1}$. Bottom row (panel e) \rev{shows how the predicted FIP bias evaluated at $s=57.5~\mathrm{Mm}$ changes} as turbulence increases.
}
    \label{fig:v_turb_investigation}
\end{figure}

\rev{The acoustic wave flux results demonstrate that turbulence in the $v_{\text{turb}}$ term of Equation~\ref{equ:v_w} strongly regulates fractionation. However, acoustic waves represent just one potential turbulence source. To explore this regulatory mechanism more generally, we now examine how any additional turbulent velocity affects predicted FIP bias.}

\rev{The effect} of weak acoustic wave flux can also be interpreted as an added turbulence in $v_{w}$. To illustrate the role of turbulence in elemental fractionation more explicitly, Figure~\ref{fig:v_turb_investigation} shows the effect of an artificially added turbulent velocity component, \(v_{\rm add}\), which could represent any chromospheric disturbance source beyond acoustic waves. This term enters the denominator of Equation~\ref{equ:FIP_bias_integration} identically to acoustic wave contributions, suppressing fractionation for a given ponderomotive acceleration. Explicitly, we now write
\begin{equation}
    v_w^2 = k_B T/m_s + v^2_\mathrm{turb} + v^2_\mathrm{wave} + v^2_{\rm add},
\end{equation}
and evaluate the predicted FIP bias. 

In these calculations, we fix the Alfv\'{e}n wave amplitude at 0.97 km s\(^{-1}\), $F_\mathrm{ac}=10^8~\mathrm{erg~cm^{-2}~s^{-1}}$, while varying $v_{\rm add}$ from 0 to 50 km s$^{-1}$. \rev{The goal is to fix the wave driver, thus isolating the effect of the added turbulence, and avoid mixing two coupled parameters.} The results demonstrate a clear inverse relationship. \rev{As expected from Equation~\ref{equ:FIP_bias_integration}, increasing turbulence systematically reduces predicted FIP bias across all elements.} At weak to moderate turbulence levels (0–10 km s\(^{-1}\); panels a--c), low-FIP elements maintain substantial fractionation, but overall fractionation becomes increasingly suppressed. All elements show reduced enhancement as turbulent velocities begin to exceed thermal speeds.

At extreme turbulence ($>30$ km s\(^{-1}\); \rev{panels} d, and e), fractionation nearly vanishes for all elements, with FIP bias approaching unity. Here, \(v_{\rm add}^2\) completely dominates the \(v_w^2\) term in Equation~\ref{equ:v_w}, overwhelming both thermal speeds and slow-mode wave contributions. The integral becomes essentially element-independent, eliminating both FIP-based and mass-based fractionation.

This effect can perhaps be understood in a more straightforward way. When chromospheric plasma experiences strong turbulent motions, the relative effect of ponderomotive acceleration diminishes because ions and neutrals are both scattered more frequently, reducing the differential transport that produces fractionation.

These turbulence levels are observationally motivated by flare nonthermal velocities inferred from EIS and IRIS spectroscopy, typically of order $10-30~\mathrm{km~s^{-1}}$ in pre-flare and impulsive phases. Pre-flare and impulsive-phase spectra routinely show increased nonthermal velocity measured with Hinode/Extreme-ultraviolet Imaging Spectrometer (EIS) and Interface Region Imaging Spectrograph (IRIS), suggesting substantial turbulence $(>10~\rm km~s^{-1})$ in the corona and chromosphere before the main energy release \citep[e.g.,][]{Jeffrey2018SciA....4.2794J,Woods2021ApJ...922..137W, To2025ApJ...993..102T}. In the context of Figure~\ref{fig:v_turb_investigation}, these observed values correspond to turbulent conditions between panels (c) and (d), as shown in panel (e), where the additional turbulent term $v_{\rm add}$ becomes comparable to or exceeds the thermal speed. Under such conditions, the predicted FIP bias is strongly suppressed. As turbulence intensifies during the pre-flare build-up and impulsive phases, the effective wave speed $v_w$ in Equation~(\ref{equ:v_w}) increases, decreasing the ponderomotive acceleration and driving the elemental fractionation toward unity, \rev{potentially} contributing to the decreasing FIP bias seen in~\citep[][]{Warren2014ApJ...786L...2W,Mondal2021ApJ...920....4M,To2024A&A...691A..95T,Suarez2023ApJ...957...14S, Brooks2024ApJ...962..105B, Telikicherla2024ApJ...966..198T, Ng2025ApJ...979...86N}.

\section{Discussion and Conclusions}

\rev{In this paper, we have shown that the ponderomotive force model predicts similar fractionation patterns under dynamically heated chromospheres, while acoustic flux and turbulence strongly regulate the final predicted FIP bias.} Using \texttt{HYDRAD} simulations combined with the new \texttt{FIPpy} code, our key findings are:
\begin{enumerate}
    \item \textbf{Chromospheric structure matters.} Heated active region chromospheres produce slightly different FIP bias patterns from quiet Sun VAL-C atmospheres \rev{(Figures~\ref{fig:0s},~\ref{fig:700s}, and \ref{fig:by_element_fip_bias_comparison})}.
    \item \textbf{Mass-dependent fractionation emerges at low acoustic wave flux.} When acoustic wave flux drops below $\sim$$5\times10^6$ erg cm$^{-2}$ s$^{-1}$, thermal velocities $v_{\rm th}$ dominate the effective speed $v_w$ in Equation~\ref{equ:v_w} (Figure~\ref{fig:integral_investigation}). Since fractionation depends on $1/v_w^2$, heavier elements experience stronger enhancement regardless of ionisation potential. This produces a counterintuitive pattern, with the FIP bias of Fe (56 amu, 7.90 eV FIP) exceeding Ca (40 amu, 6.11 eV FIP), while the high-FIP Ar (40 amu, 15.76 eV FIP) shows enhancement comparable to low-FIP Si and Mg (Figure~\ref{fig:acoustic_investigation}e). This possibly brings some alignment with recent solar wind studies, where a mass-dependent fractionation was observed~\cite{Alterman2025A&A...694A.265A,Alterman2025A&A...700A..23A}.
    \item \textbf{Turbulence universally regulates fractionation.} By adding an artificial turbulence to the ponderomotive force equation, we show that any extra turbulent velocity source suppresses fractionation by increasing $v_{\rm turb}$ in Equation~\ref{equ:v_w} (Figure~\ref{fig:v_turb_investigation}). When turbulent velocities exceed thermal speeds, the $v_w^2$ term becomes element-independent, eliminating both FIP-based and mass-based fractionation. In the extreme case (50 km s$^{-1}$), FIP bias approaches unity for all elements. 
\end{enumerate}

Our model helps explain several puzzling observations:

\textbf{(a) Temporal FIP bias variations during flares.} Flares provide an especially compelling test case for this interplay between fractionation and turbulence. Soft X-ray measurements of flares reveal a characteristic evolution: plasma composition begins at coronal values (FIP bias $\sim 3$--$4$), drops toward photospheric abundances during the impulsive phase, then recovers to coronal values during the decay phase~\cite{Mondal2021ApJ...920....4M, Nama2023SoPh..298...55N, Ng2024ApJ...972..123N}. The standard interpretation attributes these dips to chromospheric evaporation injecting fresh, unfractionated material into coronal loops~\cite{To2024A&A...691A..95T,Benavitz2025ApJ...992....4B,Reep2025arXiv250925695R}. Our results suggest turbulence plays a complementary role: as non-thermal velocities increase to tens of km~s$^{-1}$ during the pre-flare and impulsive phases~\cite{To2025ApJ...993..102T}, turbulent motions dominate the effective speed $v_w$ in Equation~(\ref{equ:v_w}), weakening the ponderomotive force contribution and driving FIP bias toward unity~\citep[seen in][]{Telikicherla2024ApJ...966..198T}. Once turbulence subsides in the decay phase, FIP bias naturally recovers to pre-flare coronal values. Thus, beyond plasma flows, flare-associated turbulence can intrinsically modify fractionation patterns.

\textbf{(b) Enhanced Fe/S FIP bias relative to Ca/Ar during flares.} Enhanced Fe/S FIP bias relative to Ca/Ar has been measured during flares~\cite{To2024A&A...691A..95T}. A delicate balance between Fe's high mass, acoustic wave flux, and total turbulence in the transition region could motivate a highly fractionated Fe, while keeping a modest Ca fractionation. 

\textbf{(c) Anomalous S and Ar behaviour in active regions.} \rev{Anomalous S and Ar behaviour was observed in active regions~\cite{To2021ApJ...911...86T,Mihailescu2023ApJ...959...72M,To2023ApJ...948..121T}}, where S is measured to be fractionated. Both elements have substantial masses (S: 32~amu, Ar: 40~amu) enabling fractionation when the overall turbulence is weak compared to the Alfvén wave amplitude within a loop. 

\textbf{(d) Spatial FIP bias variations within active regions.} Spatial FIP bias variations within individual active regions~\cite{Mihailescu2022ApJ...933..245M,Baker2024ApJ...970...39B} naturally arise if local differences in chromospheric temperature, density, and turbulence produce corresponding abundance gradients along and across loops.

These findings suggest that variations in FIP bias may reflect chromospheric heating history and turbulence levels rather than fundamental differences in fractionation mechanisms. Coronal abundances encode integrated information about chromospheric processes; decoding this requires understanding how dynamic atmospheres alter the predictions from static models.

\subsection{Limitations and future directions}

Our approach treats fractionation as post-processing applied to individual \texttt{HYDRAD} snapshots, and therefore misses the instantaneous response of the elemental ion fractions during fast events. \rev{For the fractionation calculation, the ion fractions of the relevant elements are approximated in post-processing using the hybrid Saha--CHIANTI prescription. However, the \texttt{HYDRAD}-modelled atmosphere does not assume equilibrium ionisation for hydrogen, which should provide a more realistic thermodynamic background for the post-processing. Element-specific non-equilibrium ionisation effects during the most impulsive phases are nevertheless not captured. We note that the fractionation integral is concentrated in the dense lower chromosphere and transition region where ionisation timescales are short, and that fractionation is only sensitive to the total ionised fraction rather than the detailed charge state distribution, which mitigates the impact of this approximation.} A fully self-consistent treatment would incorporate ponderomotive acceleration directly into the momentum equations, enabling dynamic feedback between fractionation and atmospheric evolution. Additionally, we launch monochromatic Alfv\'en waves rather than realistic broadband spectra and employ a simplified 1D magnetic field structure.

Despite these limitations, the \texttt{FIPpy} code bridges the critical gap between static atmospheric models and dynamic solar reality. Future applications include: investigating specific events using observed heating profiles as inputs, systematic exploration of parameter space (acoustic flux, loop length, magnetic field strength, heating rates), testing predictions against coordinated chromospheric and coronal observations, and identifying observable diagnostics that predict which fractionation regime (FIP-dominated versus mass-dominated) operates in different regions.

Beyond the solar case, an important avenue for future development is the extension of this code to other stars. Stellar coronal surveys reveal a wide diversity of abundance patterns, including strong solar-like FIP effects, 
inverse-FIP effects in active M dwarfs, and mass-dependent trends in young, rapidly rotating stars that remain poorly explained within a unified physical picture~\cite{Wood2010ApJ...717.1279W,Testa2015RSPTA.37340259T,Seli2022A&A...659A...3S,Chebly2025A&A...695A.165C,Didel2025AJ....169...49D,Kurihara2025_stars}. Extending \texttt{FIPpy} into stellar parameter space would therefore enable predictive modelling of the FIP/inverse-FIP boundary as a function of stellar type and activity level. 

This work demonstrates that chromospheric dynamics play a crucial role in determining coronal composition. The ponderomotive force model, when applied to realistic time-evolving atmospheres rather than static structures, reveals richer behaviour. Understanding these effects is essential for using coronal abundances as diagnostics of the coupled chromosphere-corona system.

\vskip6pt

\section{Software}
This paper made use of several open-source packages including 
\texttt{Astropy v7.1.1}~\cite{AstropyCollaboration2022ApJ...935..167A}, 
\texttt{matplotlib v3.9.2}~\cite{Hunter2007CSE.....9...90H}, 
\texttt{numpy v2.0.1}~\cite{Harris2020Natur.585..357H}, 
\texttt{scipy v1.16.3}~\cite{Virtanen2020NatMe..17..261V}, \texttt{Jupyter Notebook v7.2.1} \cite{jupyter2007CSE.....9c..21P,kluyver2016jupyter}, \texttt{EBTEL\footnote{\url{https://github.com/rice-solar-physics/EBTEL}}}~\cite{Klimchuk2012ascl.soft03007K, Cargill2012ApJ...752..161C,Cargill2012ApJ...758....5C}, and \texttt{HYDRAD}\footnote{\url{https://github.com/rice-solar-physics/HYDRAD}}~\cite{Bradshaw2003A&A...401..699B, Bradshaw2013ApJ...770...12B}. \rev{\texttt{EBTEL} was used for estimates of the heating parameters used to guide the \texttt{HYDRAD} setup.}
The code developed for this work, \texttt{FIPpy}, is open-source and  available at \url{https://github.com/andyto1234/FIPpy}~\cite{to_2026_fippy}.

\enlargethispage{20pt}

\ack{\rev{We thank the referees for their insightful comments that improved the manuscript.} A.S.H.T. and A.J.F. acknowledge support through the European Space Agency (ESA) Research Fellowship Programme in Space Science. A.S.H.T. also thanks the ESA faculty research grant that supported the collaboration visit to the Naval Research Laboratory and University of Hawaii that cultivated this study. }












\bibliographystyle{RS}+ 
\bibliography{bib}  

\end{document}